\documentclass{article}

\usepackage{microtype}
\usepackage{graphicx}
\usepackage{subfigure}
\usepackage{booktabs} 

\usepackage{hyperref}

\usepackage[accepted]{icml2025}

\usepackage{amsmath}
\usepackage{amssymb}
\usepackage{mathtools}
\usepackage{amsthm}

\usepackage[capitalize,noabbrev]{cleveref}

\theoremstyle{plain}

\theoremstyle{definition}

\theoremstyle{remark}

\usepackage[textsize=tiny]{todonotes}

\icmltitlerunning{Submission and Formatting Instructions for ICML 2025}

\begin{document}

\twocolumn[
\icmltitle{Improving Myocardial Infarction Detection via Synthetic ECG Pretraining}



\icmlsetsymbol{equal}{*}

\begin{icmlauthorlist}
\icmlauthor{Lachin Naghashyar}{equal,yyy}

\end{icmlauthorlist}

\icmlaffiliation{yyy}{Department of Computer Science, University of Oxford, Oxford, UK}

\icmlcorrespondingauthor{Lachin Naghashyar}{lachin.naghashyar@cs.ox.ac.uk}

\icmlkeywords{Machine Learning, ICML}

\vskip 0.3in
]



\printAffiliationsAndNotice{\icmlEqualContribution} 

\begin{abstract}
Myocardial infarction is a major cause of death globally, and accurate early diagnosis from electrocardiograms (ECGs) remains a clinical priority. Deep learning models have shown promise for automated ECG interpretation, but require large amounts of labeled data, which are often scarce in practice.  We propose a physiology-aware pipeline that (i) synthesizes 12-lead ECGs with tunable MI morphology and realistic noise, and (ii) pre-trains recurrent and transformer classifiers with self-supervised masked-autoencoding plus a joint reconstruction–classification objective. We validate the realism of synthetic ECGs via statistical and visual analysis, confirming that key morphological features are preserved. Pretraining on synthetic data consistently improved classification performance, particularly in low-data settings, with AUC gains of up to 4 percentage points. These results show that controlled synthetic ECGs can help improve MI detection when real clinical data is limited.

\end{abstract}

\section{Introduction}

Myocardial infarction (MI) remains a leading cause of morbidity and mortality worldwide, with early diagnosis via electrocardiogram (ECG) analysis being critical for effective intervention. Machine learning approaches have shown promise for automated MI detection from ECG signals; however, they typically require large quantities of labeled clinical data, which are challenging to obtain due to privacy concerns, patient variability, and expert annotation requirements. This makes it difficult to develop robust, generalizable models, particularly for rare or complex pathological presentations.

To address this challenge, researchers have explored the generation of synthetic ECGs to augment training datasets. Traditional simulation-based methods~\cite{mcsharry2003dynamical} offer high signal fidelity but limited variability, while deep generative models, such as GANs~\cite{berger2023gan_review}, produce diverse outputs but often lack direct physiological control over clinically important features. In contrast, our approach explicitly models cardiac cycles, enabling precise modulation of MI-related morphology such as Q-wave amplitude and ST-segment elevation. By parameterizing waveforms directly through interpretable kernels, we can precisely control specific abnormalities like ST-elevation with clinical meaning, ensuring pathophysiological interpretability.

Meanwhile, deep learning methods, including recurrent and Transformer-based neural networks, have shown strong potential for ECG classification~\cite{ansari2023deep}, particularly when paired with self-supervised pretraining strategies like masked autoencoding~\cite{zhou2023masked,sawano2024applying}, enabling robust learning from limited labeled data and improving generalization.

In this study, we present a computational pipeline that synthesizes realistic twelve-lead ECG recordings with MI-specific alterations and leverages them to pretrain recurrent and Transformer-based neural networks. Beyond pretraining, we incorporate advanced fine-tuning strategies, including masked autoencoding, joint reconstruction-classification objectives, dynamic loss weighting, and a structured augmentation pipeline to enhance generalization and robustness.

While prior studies have independently explored synthetic ECG generation and pretraining techniques, the integration of physiologically grounded synthetic data with modern fine-tuning paradigms for MI classification remains largely underexplored. In this paper, we investigate the question of whether combining physiologically controlled synthetic ECGs with self-supervised deep learning can improve MI detection under limited data conditions. By enhancing robustness and generalization, such approaches help enable earlier and more reliable screening in prehospital or data-scarce clinical environments.

\section{Methods}
\subsection{Dataset}

We utilized the PTB-XL dataset, containing 21,837 10-second 12-lead ECG recordings from 18,885 patients, annotated with diagnostic statements following the SCP-ECG standard~\cite{wagner2020ptb}. For this study, PTB-XL was employed for fine-tuning and evaluation phases, with synthetic ECGs used during pre-training. We adopted the recommended patient-stratified splits: folds 1–8 for training, fold 9 for validation, and fold 10 for testing. Signals were resampled to 100~Hz and standardized to 1000 timesteps per lead. Numerical metadata (age, height, weight) were normalized using Min-Max scaling after mean imputation, while categorical metadata (sex) was label-encoded. Preprocessing parameters were fitted on the training set and applied consistently to validation and test sets.

\subsection{Synthetic ECG Simulation}

Synthetic twelve-lead electrocardiograms were generated at 100\,Hz by adapting \texttt{NeuroKit2}~\cite{Makowski2021neurokit} to a multilead setting with limb and precordial transfer matrices. Each virtual beat was modeled by summing Gaussian-shaped P, Q, R, S, and T kernels, with timings ($t_i$), amplitudes ($a_i$), and widths ($b_i$) sampled from class-conditioned normal distributions derived from PTB-XL statistics, ensuring realistic atrial depolarization, ventricular depolarization, and repolarization phases. Heart-rate variability was introduced by sampling RR intervals from a log-normal distribution fitted to clinical data and filtered for physiological LF/HF balance. 

For MI cases, pathology-specific alterations were introduced: Q-wave deepening, ST-segment elevation (0.1–0.3\,mV post-R-peak), T-wave inversion and scaling, and QRS broadening to simulate ischemic conduction delays. Beat-to-beat amplitude jitter, localized voltage distortions near R-peaks, and lead-wise timing shifts were added to reflect acute infarction variability. The signals were also augmented with physiological artifacts, including respiratory baseline wander, mains interference, and electromyographic noise with greater variability in MI cases. Transient motion artifacts during chest pain episodes were simulated by disturbances near R-peaks in MI traces. A nonlinear fade-in ramp was applied at the start of each record, deterministic for normal cases, and stochastic for MI. Each lead was normalized and randomly scaled to reflect inter-device calibration differences. Synthetic datasets were generated in mini-batches across diagnostic classes.

The fidelity of the simulated cohort was assessed by squared Maximum Mean Discrepancy (MMD) between real and synthetic signals using a Gaussian kernel with bandwidth selected by the median heuristic. Additional validation included Kolmogorov–Smirnov tests on flattened waveforms, lead-wise R-peak amplitudes, and basic signal features (mean, standard deviation, peak-to-peak voltage), along with power-spectral density comparisons in the clinical band. Visual inspection of randomly paired real and synthetic signals further supported verification.

\subsection{RNN-based ECG Classifier}

We developed a recurrent neural network (RNN) classifier to process 12-lead ECG signals and structured metadata. GRUs were chosen for their efficiency in modeling sequential dependencies typical of ECG signals. A bidirectional GRU encoder~\cite{cho2014learning} was applied to the signals, followed by average and max pooling over time. Pooled features were concatenated with embedded categorical (e.g., sex) and normalized numerical metadata (e.g., age, height, weight), then passed through a fully connected layer with ReLU activation, batch normalization, and dropout, followed by a final linear output layer.

Two versions of the classifier were trained: (i) a \emph{multi-label classifier} predicting superclass and subclass labels using sigmoid activation and binary cross-entropy loss, and (ii) a \emph{binary classifier} focused solely on MI detection. Although the synthetic ECGs contained only two classes (normal and MI), the multi-label setting enabled future fine-tuning on real-world multi-label data.

Training used AdamW optimization with cosine-annealed learning rates and mixed-precision gradient scaling. Early stopping was based on validation loss, and best-performing models were checkpointed. In some experiments, pretraining was conducted on synthetic ECGs to initialize the GRU encoder with pathology-aware representations, improving convergence and robustness when fine-tuning on limited real-world data.

\subsection{Transformer-based ECG Classifier}

We developed a Transformer-based model~\cite{vaswani2017attention} to classify 12-lead ECG signals for MI detection. To capture the varying morphologies of ECG waveforms, including QRS complexes, P-waves, and T-waves, input signals were first projected through a one-dimensional convolutional layer and passed through multiscale convolutional feature extraction blocks with multiple kernel sizes. The features are combined and encoded with position encodings and passed through a multi-layer Transformer encoder. The final per-sample embeddings are pooled and concatenated with demographic embeddings and numerical metadata, then passed through a fully connected classifier head to output a binary MI prediction.

To enhance performance and address limited real data availability, we performed two types of pretraining on synthetic ECGs. First, we conducted self-supervised pretraining with a masked autoencoder (MAE) objective~\cite{zhou2023masked, sawano2024applying}, randomly masking signal segments and training the model to reconstruct the missing parts. Second, we performed joint pretraining combining signal reconstruction and MI classification with a dynamic loss weighting scheme. Both approaches used augmentation strategies such as amplitude scaling, time masking, noise injection, baseline wander, and lead dropout.

During fine-tuning, the model was initialized with pretrained weights and trained on a stratified subset of real ECGs. Balanced binary cross-entropy with mild focal weighting addressed class imbalance. Cosine-annealed learning rates and gradient clipping were applied for stability. Early stopping was based on validation AUC, and models were evaluated on a held-out test set using AUROC, accuracy, sensitivity, specificity, and F1-score. Experiments used PyTorch 2.0.1, NumPy 1.24, and SciPy 1.10, with fixed random seeds and ran on an NVIDIA L4 GPU.
 \href{https://anonymous.4open.science/r/Computational-Medicine-F65F/Improving_Myocardial_Infarction_Detection_via_Synthetic_ECG_Pretraining.ipynb}{Code and reproducibility materials available here.}

\section{Results}

\subsection{Analysis of Synthetic ECG Results}

To verify the realism of the synthetic ECGs, we evaluated the distributional similarity between synthetic and real signals using the squared Maximum Mean Discrepancy (MMD\(^2\)), obtaining scores of 0.095 for healthy signals and 0.073 for MI, indicating moderate alignment. These values are similar to those reported for GAN-based ECG synthesis on PTB, which typically range from 0.01 to 0.11~\cite{berger2023gan_review}.

\begin{figure}[h]
    \centering
    \begin{subfigure}{}
        \centering
        \includegraphics[width=\linewidth]{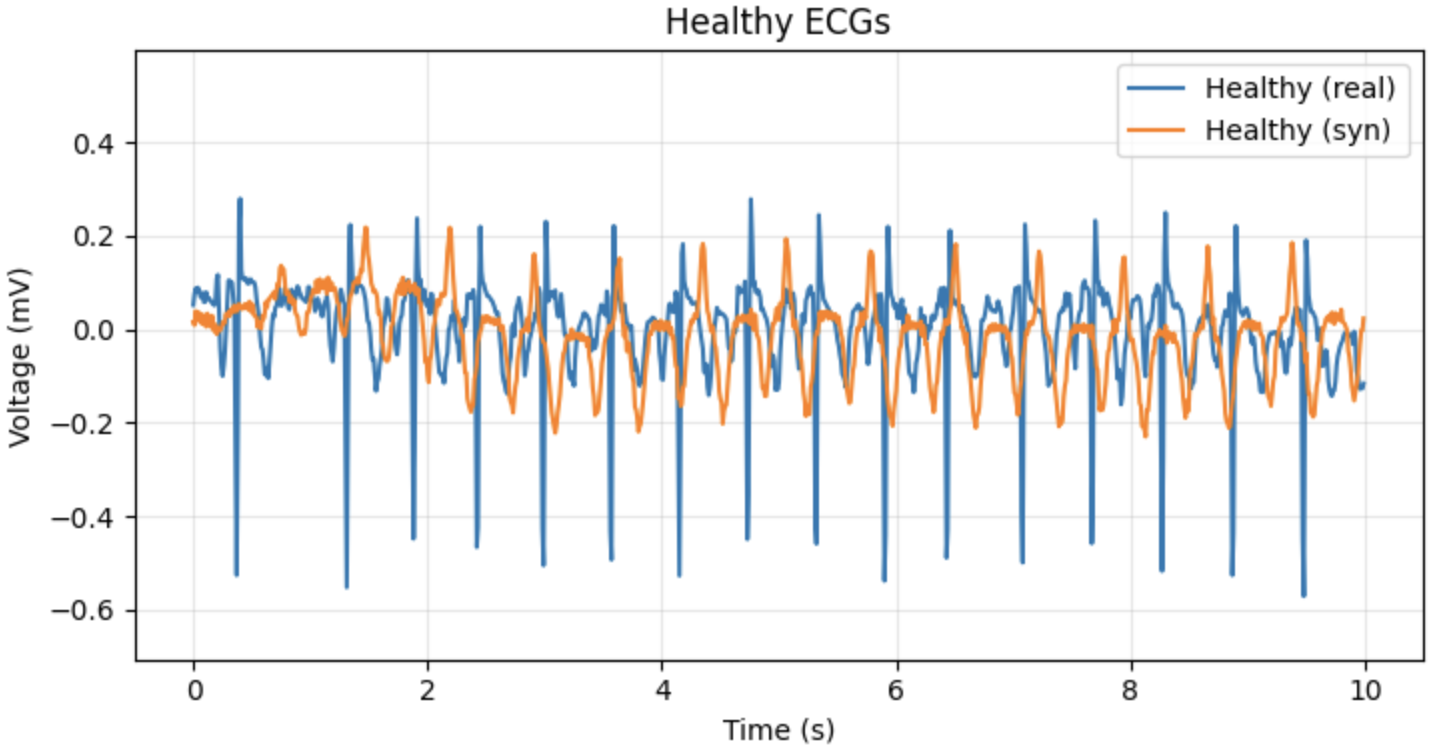}
        \caption{Healthy ECGs.}
        \label{fig:healthy_ecg}
    \end{subfigure}
    \hfill
    \begin{subfigure}{}
        \centering
        \includegraphics[width=\linewidth]{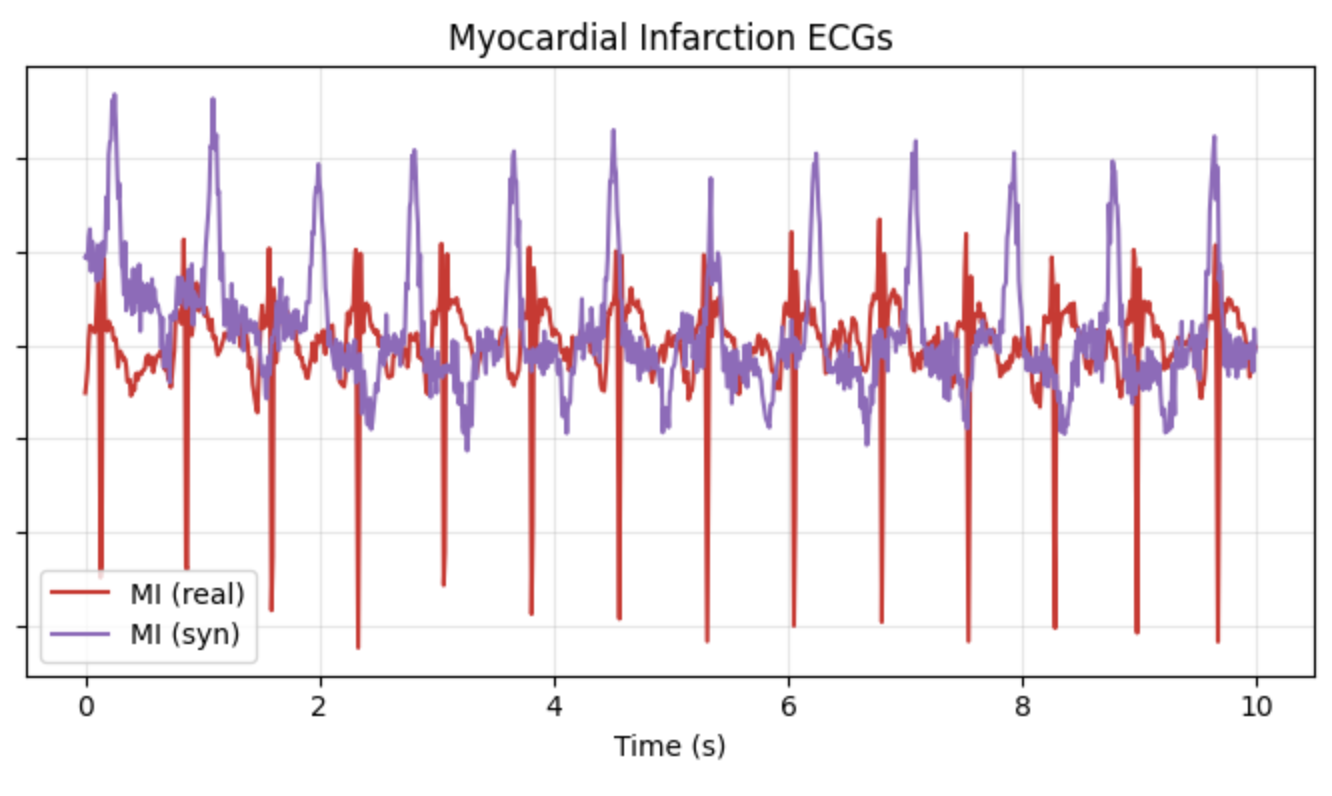}
        \caption{MI ECGs.}
        \label{fig:mi_ecg}
    \end{subfigure}
    \caption{Comparison of randomly paired real and synthetic ECG signals for healthy and MI cases.}
    \label{fig:ecg_real_syn_comparison}
\end{figure}

Additional validation with two-sample Kolmogorov--Smirnov (KS) tests showed a KS distance of 0.066 for real and 0.21 for synthetic flattened signal distributions, reflecting some residual mismatch. Per-lead KS analysis yielded an average distance of 0.177±0.031 across leads, likely due to limited variability in pathological cases. Intra-group KS distances remained low for both real and synthetic datasets (0.007–0.019), confirming strong internal consistency and coherent class-specific morphology. Qualitative verification by visual comparison of random real and synthetic traces (Figures~\ref{fig:healthy_ecg} and~\ref{fig:mi_ecg}) showed that synthetic signals captured key morphological features, more detailed plots are available in the \href{https://anonymous.4open.science/r/Computational-Medicine-F65F/Improving_Myocardial_Infarction_Detection_via_Synthetic_ECG_Pretraining.ipynb}{notebook.}

\subsection{Multi-label Classification Results}

We evaluated the impact of synthetic pretraining on a multi-label GRU-based ECG classifier. As shown in Table~\ref{tab:multilabel_auc_results}, pretraining consistently improved the averaged superclass AUC across varying amounts of real MI data. In the most real MI-scarce regime (\textit{frac} = 0), synthetic pretraining raised the AUC from 83.1\% to 86.6\%. With more real data available, both base and pretrained models achieved high AUCs (above 92\%), though pretraining still provided small but consistent gains. 

Recent studies on PTB-XL have reported superclass AUCs between 87\% and 93\%, depending on model complexity and supervision levels~\cite{strodthoff2021ptbxl}. Our pretrained GRU-based model achieved competitive performance, particularly in low-data settings, highlighting the benefit of synthetic pretraining.

\begin{table}[h]
\centering
\caption{AUROC (\%) for multi-label ECG classification across different fractions of real data.}
\label{tab:multilabel_auc_results}
\begin{tabular}{|c|c|c|c|c|}
\hline
\textbf{Model} & \textbf{Frac 0.0} & \textbf{Frac 0.1} & \textbf{Frac 0.5} & \textbf{Frac 1.0} \\
\hline
Base & 83.1 & 88.9 & 88.9 & 92.7 \\
Pretrained & 86.6 & 90.1 & 91.5 & 92.6 \\
\hline
\end{tabular}
\end{table}

\subsection{Binary Classification Results}

We evaluated the impact of synthetic pretraining on binary MI detection using GRU-based and Transformer-based models across varying synthetic-to-real data ratios. Synthetic pretraining consistently improved AUC for the GRU model, while Transformer models benefited more from self-supervised MAE and joint reconstruction-classification pretraining, showing the advantages of structure-aware pretraining on synthetic data. Prior work on PTB-XL~\cite{strodthoff2021ptbxl} has reported binary MI detection AUCs ranging from 85\% to 98\%~\cite{sun2025multi}, consistent with the performance observed in our experiments. For the best-performing model, bootstrap resampling (n=1000) yielded an estimated AUC of 89.3\% (95\% CI: 87.8\%–90.7\%), supporting the robustness of the performance. These results support the hypothesis that synthetic data with self-supervised pretraining improves MI detection under limited data.

\begin{table}[h]
\centering
\caption{AUROC (\%) for binary MI detection with different models and synthetic-to-real data ratios.}
\label{tab:binary_auc_results}
\renewcommand{\arraystretch}{1.2}
\begin{tabular}{|c|c|c|c|c|}
\hline
\textbf{Model} & \textbf{1.0} & \textbf{0.5} & \textbf{0.1} & \textbf{0.05} \\
\hline
GRU Base & 80.6 & 82.8 & 89.9 & 92.12 \\
GRU Pretrained & 81.7 & 83.4 & 89.9 & 91.54 \\
\hline
Transformer Base & 78.8 & 78.6 & 83.7 & 86.42 \\
Transformer MAE & 82.0 & 82.2 & 87.1 & 89.23 \\
Transformer Joint & 82.0 & 82.3 & 86.9 & 89.52 \\
\hline
\end{tabular}
\end{table}

\subsection{Verification, Validation, and Sensitivity Analysis}

\textbf{Verification.}
All simulation procedures were functionally, manually, and visually verified to ensure morphological consistency with clinical expectations. Random inspections confirmed that key features (e.g., ST-segment elevation) matched clinical expectations. Classifier training pipelines were verified by monitoring convergence plots of loss and AUC on training and validation sets.

\textbf{Validation.}
The realism of synthetic ECGs was validated through statistical comparisons with real PTB-XL data, achieving moderate alignment based on MMD and KS metrics. Additional qualitative validation confirmed that key waveform morphologies were preserved (Figures~\ref{fig:healthy_ecg}, \ref{fig:mi_ecg}). Classifier models were evaluated on held-out real data using AUC, accuracy, sensitivity, specificity, and F1-score. Calibration plots were also used to assess the alignment between predicted probabilities and observed outcomes, supporting probabilistic interpretability and model reliability.

\textbf{Sensitivity Analysis.}
Sensitivity of model performance was evaluated across varying real-to-synthetic data ratios and different pretraining strategies. In multi-label classification, synthetic pretraining showed larger benefits under data-scarce regimes (e.g., \textit{frac}=0). In binary MI detection, Transformer models exhibited greater sensitivity to self-supervised MAE and joint pretraining strategies, particularly under low real data availability. Sensitivity to hyperparameters (e.g., learning rate schedules, augmentation strength) was qualitatively monitored, although no formal hyperparameter grid search was performed. Further uncertainty quantification was conducted by applying bootstrap resampling (n=1000) to estimate confidence intervals for the test set AUC of the best-performing model.

\section{Conclusions}

\textbf{Key findings and contributions.}
We introduced a fully reproducible pipeline that (i) synthesizes physiologically-controlled 12-lead ECGs with tunable MI morphology, (ii) uses those signals for self-supervised pre-training of GRU and Transformer classifiers, and (iii) achieves state-of-the-art performance when only a small fraction of real PTB-XL data are available. Synthetic pre-training raised multi-label AUC up to 4 percent in the zero-real-MI regime and yielded a best binary-MI AUC of 91.54\%. These results confirm that physiologically grounded augmentation, coupled with masked-autoencoding or joint objectives, can improve MI detection under data-scarce conditions.

\textbf{Strengths and limitations.}
Strengths include explicit physiological control of waveform morphologies (improving interpretability), comprehensive V\&V with MMD/KS statistics and bootstrap CIs, and modular code that easily extends to other arrhythmias. Limitations include the smoother beat-to-beat variability of simulated ECGs compared to real clinical recordings, the use of a single public dataset without cross-hospital validation, and a heuristic hyperparameter search, which may not yield globally optimal performance. Additionally, while the synthetic data captured common MI presentations, rarer patterns such as posterior infarction were not explicitly modeled.

\textbf{Future work and pathway to impact.}
The next steps will focus on (i) enriching the simulator with posterior-/inferior-wall and multi-pathology combinations, (ii) validating the pipeline prospectively on an ambulance or wearable ECG streams, and (iii) embedding the lightweight GRU model on edge devices for real-time pre-hospital triage. Another future direction is to compare our waveform-based synthesis with anatomy-driven ECG simulations from biophysical heart models~\cite{riebel2024insilico}. To support real-world use, the model could be fine-tuned on local data using privacy-preserving methods like federated learning. A clinical study could then compare model-assisted triage to standard care to see if it speeds up treatment. Ultimately, this approach could help reduce treatment delays and improve outcomes in low-resource settings.

\bibliographystyle{icml2025}  
\bibliography{example_paper}  

\begin{thebibliography}{12}
\providecommand{\natexlab}[1]{#1}
\providecommand{\url}[1]{\texttt{#1}}
\expandafter\ifx\csname urlstyle\endcsname\relax
  \providecommand{\doi}[1]{doi: #1}\else
  \providecommand{\doi}{doi: \begingroup \urlstyle{rm}\Url}\fi

\bibitem[Ansari et~al.(2023)Ansari, Mourad, and Qaraqe]{ansari2023deep}
Ansari, Y., Mourad, O., and Qaraqe, K.
\newblock Deep learning for ecg arrhythmia detection and classification: an overview of progress for period 2017--2023.
\newblock \emph{Frontiers in Physiology}, 2023.
\newblock \doi{10.3389/fphys.2023.1246746}.

\bibitem[Berger et~al.(2023)Berger, Haberbusch, and Moscato]{berger2023gan_review}
Berger, L., Haberbusch, M., and Moscato, F.
\newblock Generative adversarial networks in electrocardiogram synthesis: Recent developments and challenges.
\newblock \emph{Artificial Intelligence in Medicine}, 2023.
\newblock \doi{10.1016/j.artmed.2023.102632}.

\bibitem[Cho et~al.(2014)Cho, Van~Merriënboer, Gulcehre, Bahdanau, Bougares, Schwenk, and Bengio]{cho2014learning}
Cho, K., Van~Merriënboer, B., Gulcehre, C., Bahdanau, D., Bougares, F., Schwenk, H., and Bengio, Y.
\newblock Learning phrase representations using rnn encoder-decoder for statistical machine translation.
\newblock \emph{arXiv preprint arXiv:1406.1078}, 2014.

\bibitem[Makowski et~al.(2021)Makowski, Pham, Lau, Brammer, Lespinasse, Pham, Schölzel, and Chen]{Makowski2021neurokit}
Makowski, D., Pham, T., Lau, Z.~J., Brammer, J.~C., Lespinasse, F., Pham, H., Schölzel, C., and Chen, S. H.~A.
\newblock {NeuroKit}2: A python toolbox for neurophysiological signal processing.
\newblock \emph{Behavior Research Methods}, 2021.
\newblock \doi{10.3758/s13428-020-01516-y}.

\bibitem[McSharry et~al.(2003)McSharry, Clifford, and Tarassenko]{mcsharry2003dynamical}
McSharry, P.~E., Clifford, G.~D., and Tarassenko, L.
\newblock A dynamical model for generating synthetic electrocardiogram signals.
\newblock \emph{IEEE Transactions on Biomedical Engineering}, 2003.

\bibitem[Riebel et~al.(2024)Riebel, Camps, Arantes~Berg, Rodriguez, et~al.]{riebel2024insilico}
Riebel, L., Camps, J., Arantes~Berg, L., Rodriguez, B., et~al.
\newblock In silico evaluation of cell therapy in acute versus chronic infarction.
\newblock \emph{Scientific Reports}, 2024.
\newblock \doi{10.1038/s41598-024-67951-5}.

\bibitem[Sawano et~al.(2024)Sawano, Yamakawa, Ohnishi, Nakamura, and Nishiyama]{sawano2024applying}
Sawano, M., Yamakawa, T., Ohnishi, A., Nakamura, S., and Nishiyama, K.
\newblock Applying masked autoencoder-based self-supervised learning for high-capability vision transformers of electrocardiographies.
\newblock \emph{PLOS ONE}, 2024.
\newblock \doi{10.1371/journal.pone.0296245}.

\bibitem[Strodthoff et~al.(2021)Strodthoff, Wagner, Schaeffter, and Samek]{strodthoff2021ptbxl}
Strodthoff, N., Wagner, P., Schaeffter, T., and Samek, W.
\newblock Deep learning for ecg analysis: Benchmarks and insights from ptb-xl.
\newblock \emph{IEEE Journal of Biomedical and Health Informatics}, 2021.
\newblock \doi{10.1109/JBHI.2020.3037127}.

\bibitem[Sun et~al.(2025)Sun, Li, Liang, Liu, Pang, Chen, and Wang]{sun2025multi}
Sun, Q., Li, J., Liang, C., Liu, R., Pang, J., Chen, Y., and Wang, C.
\newblock A multi-lead group network for myocardial infarction detection and localization based on clinical knowledge-driven and dynamic-static feature fusion.
\newblock \emph{Expert Systems with Applications}, 2025.
\newblock \doi{10.1016/j.eswa.2025.126901}.

\bibitem[Vaswani et~al.(2017)Vaswani, Shazeer, Parmar, Uszkoreit, Jones, Gomez, Kaiser, and Polosukhin]{vaswani2017attention}
Vaswani, A., Shazeer, N., Parmar, N., Uszkoreit, J., Jones, L., Gomez, A.~N., Kaiser, {\L}., and Polosukhin, I.
\newblock Attention is all you need.
\newblock In \emph{Advances in Neural Information Processing Systems}, 2017.

\bibitem[Wagner et~al.(2020)Wagner, Strodthoff, Bousseljot, Kreiseler, Lunze, Samek, and Schaeffter]{wagner2020ptb}
Wagner, P., Strodthoff, N., Bousseljot, R.-D., Kreiseler, D., Lunze, F.~I., Samek, W., and Schaeffter, T.
\newblock Ptb-xl, a large publicly available electrocardiography dataset.
\newblock \emph{Scientific Data}, 2020.

\bibitem[Zhou et~al.(2023)Zhou, Diao, Huo, Liu, Fan, and Zhao]{zhou2023masked}
Zhou, Y., Diao, X., Huo, Y., Liu, Y., Fan, X., and Zhao, W.
\newblock Masked transformer for electrocardiogram classification.
\newblock \emph{arXiv preprint:2309.07136}, 2023.

\end{thebibliography}

\end{document}